\renewcommand{\v}[1]{{\bf #1}}

\newcommand{\s}{{\sigma}}

\def\be{\begin{eqnarray}}
\def\ee{\end{eqnarray}}
\newcommand{\nn}{\nonumber\\}

\newcommand{\Eq}[1]{Eq.~(\ref{#1})}
\newcommand{\e}{\epsilon}
\newcommand{\p}{\partial}
\newcommand{\ua}{\uparrow}
\newcommand{\da}{\downarrow}
\newcommand{\ra}{\rightarrow}

\documentclass[prl,twocolumn,showpacs,preprintnumbers,amsmath,amssymb]{revtex4}
\usepackage{graphicx}% Include figure files
\usepackage{dcolumn}% Align table columns on decimal point
\usepackage{bm}% bold math
\usepackage{wasysym}% shapes
\begin{document}
%\preprint{NSF-KITP-06-34}
\title{An Index Theorem for the Majorana Zero Modes in Chiral
P-Wave Superconductors
%The Majorana Zero Mode in the Vortex Core of a $p_x +ip_y$ Superconductor: an Index Theorem \\
\vskip 1mm
       %\small{\rm{(KITP Preprint NSF-KITP-05-31)}}
       %\small{$[$ Phys. Rev. B {\bm 63}, 174428 (2001) $]$
}
\author{Sumanta Tewari$^1$, S. Das Sarma$^1$, Dung-Hai Lee$^2$}
\affiliation{$^{1}$Condensed Matter Theory Center, Department of
Physics, University of Maryland, College Park, MD 20742\\
$^{2}$Department of Physics, University of California at Berkeley,
Berkeley, CA 94720}
%             $^{2}$Center for Condensed Matter Theory, Indian Institute
%of Science, Bangalore 560012, India\\
%         $^{3}$Department of
%Physics, Massachusetts Institute of Technology, Cambridge,
%Massachusetts 02139}
\date{\today%\\ ..KITP Preprint NSF-KITP-05-31}
}
\begin{abstract}
We show that the Majorana fermion zero modes in the cores of odd
winding number vortices of a 2D $p_x+ip_y$-paired superconductor is
due to an index theorem. This theorem is analogous to that proven by
Jackiw and Rebbi for the existence of localized Dirac fermion zero
modes on the mass domain walls of a 1D Dirac theory. The important
difference is that, in our case, the theorem is proven for a two
component fermion theory where the first and second components are
related by parity reversal and hermitian conjugation.
\end{abstract}


\maketitle

The odd winding number vortices in a spinless (spin-polarized)
$p_x+ip_y$-paired superconductor or superfluid (for brevity we
shall use ``superconductor'' to denote both in the rest of the
paper) trap zero-energy bound states in the cores. The
second-quantized operators creating these zero energy excitations
are the self-hermitian, Majorana fermion operators. This property
endows the vortices with non-Abelian statistics
\cite{Read,Ivanov,Stern}, which means that under braiding of any
two vortices the total wavefunction transforms as a vector in a
finite dimensional Hilbert space. Because of the non-Abelian
statistics, this type of superconductor has been proposed to
support topological quantum computation \cite{Ivanov,Tewari,Half}.
For example it has been shown that the Pfaffian $\nu=\frac{5}{2}$
fractional quantum Hall state\cite{Read,Moore}  is a spinless
$p_x+ip_y$ superconductor of the ``composite
fermions''\cite{Read}. The braiding of non-Abelian quasiparticles
in such a state\cite{Nayak} can, in principle, be exploited to
build a quantum computer that is immune to environmental errors
\cite{Kitaev}.

The root of the non-Abelian statistics is the presence of Majorana
fermion zero modes in the vortex cores. Such zero modes were first
proposed to exist in the vortex cores of the $p_x+ip_y$-paired
superfluid A-phase of Helium 3 in Ref.~\onlinecite{Kopnin}.
Traditionally this peculiar kind of vortex core state was
demonstrated by explicitly solving the Bogoliubov-de-Gennes (BdG)
equations\cite{Read,Stone1,Stone2,Tewari}. Interestingly, the
zero-energy solution is found only in the case when the winding
number of the vortex is an odd integer (for even winding number
vortices, there is no zero energy mode).

The zero mode in the vortex core discussed above is very analogous
to the zero energy domain wall states of polyacetylene\cite{Su}.
In that case, by writing down an ansatz for the dimerization-order
-parameter profile, one can also demonstrate the existence of the
zero energy domain wall solution by explicitly solving the
mean-field equations\cite{Maki}. Quite satisfactorily, this domain
wall state was shown to be a condensed matter realization of the
zero mode associated with the mass solitons of a 1D Dirac theory
investigated earlier by Jackiw and Rebbi\cite{Jackiw1,Jackiw2}.
The Jackiw and Rebbi soliton solution is a simple example of an
index theorem where fermionic zero modes can be used to count the
topological defects (or magnetic flux quanta) of a background
order parameter (or magnetic field).

In this paper, we ask what the analogous index theorem is for the
vortex Majorana fermion zero modes of a $p_x+ip_y$ superconductor.
We proceed by mapping the 2D vortex problem on an effective 1D
problem by performing angular momentum decomposition with respect
to the center of the vortex. In this way, we can show that for odd
winding number vortices there exists a {\it unique angular
momentum channel} in which the following Hamiltonian describes the
low energy quasiparticle excitations: \be H_M=\int
dx\Big[-iv_F\chi^{\dagger}\sigma_z\p_x\chi
+m(x)\chi^{\dagger}\sigma_x\chi\Big].\label{h}\ee In \Eq{h},
$m(-x)=-m(x)$ is a spatially varying mass term that changes sign
at $x=0$ (the location of the domain wall), and
$\chi^{\dagger}(x)$ is a two component field given by $
\chi^{\dagger}(x)=\begin{pmatrix}c^{\dagger}(x),&c(-x)\end{pmatrix},$
with $c(x)$ being a spinless fermion field. Note the important
difference of \Eq{h} with the  Dirac theory,  \be H_{D}=\int
dx\Big[-iv_F\psi^{\dagger}\sigma_z\p_x\psi
+m(x)\psi^{\dagger}\sigma_x\psi\Big],\label{h2}\ee where $
\psi^{\dagger}(x)=\begin{pmatrix}f^{\dagger}_1(x),&
f^{\dagger}_2(x)\end{pmatrix}$ with $f_{1,2}(x)$ being two {\it
independent} fermion fields. It is because of such difference, the
zero energy quasiparticles localized on the mass domain walls of
\Eq{h} are Majorana fermions, while those localized on the domain
walls of \Eq{h2} are ordinary fermions.  As we will show later,
for even winding number vortices, such angular momentum channel
does not exist.
\\

We first briefly review the derivation of the Jackiw-Rebbi zero
mode for \Eq{h2}. The quasiparticle operator $$ q^{\dagger}=\int
dx~ [\phi_1(x)f^{\dagger}_1(x)+\phi_2(x) f^{\dagger}_2(x)]$$
satisfies $[H,q^{\dagger}]=\epsilon q^{\dagger}$. That implies the
following Dirac equation for the two component wavefunction
$\phi^{\rm{T}}(x)=(\phi_1(x),\phi_2(x))$: \be
-iv_F\sigma_z\partial_x\phi(x)+\sigma_x m(x)\phi(x)=\epsilon
\phi(x). \label{wa} \ee  First we note that because $\s_y$
anticommutes with $\s_x$ and $\s_z$, if $\phi(x)$ is an
eigenfunction with eigenvalue $\epsilon$, $\sigma_y\phi(x)$ is
also an eigenfunction with eigenvalue $-\epsilon$. As a result,
the $\epsilon=0$ solutions of \Eq{wa} can be made a simultaneous
eigenstate of $\s_y$. Let $\phi_0(x)$ denote such a solution and
$\s_y\phi_0(x)=\lambda\phi_0(x)$. Set $\e=0$ and left-multiplying
\Eq{wa} by $i\s_z$ we obtain \be \p_x\phi_0(x)={\lambda\over v_F}
m(x)\phi_0(x)\nonumber,\ee which implies \be
\phi_0(x)=e^{{\lambda\over v_F}\int_0^x
m(y)dy}\phi_0(0).\label{zero}\ee For $m(x)=\pm {\rm
sign}(x)|m(x)|$, \Eq{zero} is normalizable for $\lambda=\mp 1$. In
this way we have proven that for each sign change of $m(x)$ there
is a single zero energy mode.

Now let us consider a uniform (i.e. with no vortices) 2D
$p_x+ip_y$ superconductor. The fermionic mean-field Hamiltonian is
given by $K+H_{0\rm {P}}$, where \be &&K=\sum_{\bf{k}}\xi_k
c^{\dagger}_{\bf{k}}c_{\bf{k}}\nn&&
H_{0\rm{P}}=-\Delta_0\sum_{\bf{k}}
(k_x+ik_y)c^{\dagger}_{\bf{k}}c^{\dagger}_{-\bf{k}}+\rm{h.c.}.\label{nv}
\ee Here $K$ is the kinetic energy term, $H_{0\rm{P}}$ is the
pairing term and $\xi_{k}=k^2/2m-\epsilon_{\rm{F}}$ with
$\epsilon_{\rm{F}}$ the Fermi energy.

With the purpose of treating the electronic state of a single
vortex in mind, we change to a new representation where the
fermion operators are expanded in angular momentum channels, $$
c_{\bf{k}}={1\over \sqrt{2\pi
k}}\sum_{m=-\infty}^{\infty}c_{m,k}e^{im\theta_{\bf{k}}},$$ with
$m$ an integer. The commutation relation $\{c_{\v
k},c^{\dagger}_{\v p}\}=\delta^2(\v k-\v p)$ implies $$
\{c_{mk},c^{\dagger}_{np}\}=\delta_{m,n}\delta(k-p).$$ Inserting
them in \Eq{nv}, doing the $\theta_{\bf{k}}$ integral, and
linearizing $\xi_k$ around $k_{\rm{F}}$, we get, \be
K=\frac{1}{(2\pi)^2}\sum_{m}\int_{-\Lambda}^{\Lambda}dq~(v_{\rm{F}}q)~c^{\dagger}_{m,q}c_{m,q},
\label{h0} \ee where $q=k-k_F$, $v_F$ is the Fermi velocity, and
$\Lambda$ is a momentum cut-off. So the kinetic energy term
separates into uncoupled angular momentum channels indexed by the
integer angular momentum $m$. Using similar manipulations to
decompose the pairing term, we find, using $\theta_{-\bf{k}}=\pi +
\theta_{\bf{k}}$ and $k_x+ik_y = |k|e^{i\theta_{\bf{k}}}$, \be
H_{0\rm{P}}=\frac{\Delta_0k_F
}{2\pi^2}\sum_{m}\int_{-\Lambda}^{\Lambda}dq
\cos(m\pi)c^{\dagger}_{m,q}c^{\dagger}_{1-m,q}+\rm{h.c.}
\label{hp} \ee Putting \Eq{h0} and \Eq{hp} together we have, for
each pair of $m$ and $1-m$, the following massive Dirac theory \be
H_{m}=\int dx\Big[-iv^{\prime}_F\psi_m^{\dagger}\sigma_z\p_x\psi_m
+m_0\psi_m^{\dagger}\sigma_x\psi_m\Big].\nonumber\ee In the above,
$\psi^{\dagger}_m(x)$ is the Fourier transform of
$(c^{\dagger}_{m,q},c_{1-m,q})$,
$v^{\prime}_{F}=\frac{v_F}{4\pi^2}$ and $m_0=\frac{k_F\Delta_0
\cos(m\pi)}{2\pi^2}$. Consequently, all fermionic quasiparticle
excitations are gapped.

Next we consider the fermionic Hamiltonian for a single winding
number vortex located at the origin. The kinetic energy part of the
Hamiltonian remains the same as in \Eq{h0}. Let us now consider the
pairing term.  In order to describe the spatial dependence of the
superconducting order parameter, we start with the real space
description,
\begin{equation}
H_{1\rm{P}}=-\Delta_0\int d^2R\int d^2r
e^{i\theta_{\bf{R}}}h(R)g({\bf{r}})c^{\dagger}_{{\bf{R}}+{\bf{r}}}c^{\dagger}_{{\bf{R}}-{\bf{r}}}+
\rm{h.c.}. \label{Pairingvortex}
\end{equation}
Here, $\bf{R}$ and $\bf{r}$ are the center-of-mass and the
relative coordinates of the Cooper pair, respectively. $h(R)$ and
$\theta_{\v R}$ are the amplitude and the phase of the
superconducting order parameter, and $g(\bf{r})$ is the Fourier
transform of ($p_x+ip_y$). In the vortex core,
$h(R)\sim(1-e^{-\frac{R}{\xi}})$ with $\xi$ the coherence length.
Substituting
$c^{\dagger}_{\bf{R}\pm\bf{r}}=2\pi\sum_{\bf{k}}c^{\dagger}_{\bf{k}}e^{i\bf{k}.(\bf{R}\pm\bf{r})}$
in \Eq{Pairingvortex}, we end up with two spatial integrals, $
g({\bf{k}}-{\bf{p}})=\int d^2r
g({\bf{r}})e^{i({\bf{k}}-{\bf{p}}).{\bf{r}}}=(k_x-p_x)+i(k_y-p_y)$
%\label{Eq:Integralg
and \begin{equation} I({\bf{k}}+{\bf{p}})=\int d^2R
e^{i\theta_{\bf{R}}}h(R)e^{i({\bf{k}}+{\bf{p}}).{\bf{R}}}.\label{I}\end{equation}
%\label{I}\ee
In order to evaluate $I({\bf k}+{\bf p})$, we first note that, $
I(R_{\theta}({\bf{k}}+{\bf{p}}))=e^{i\theta}I({\bf{k}}+{\bf{p}}$,
where $R_{\theta}$ is the operator that rotates
$({\bf{k}}+{\bf{p}})$ by an angle $\theta$ in the momentum space.
It follows that,
$I({\bf{k}}+{\bf{p}})=e^{i\theta_{{\bf{k}}+{\bf{p}}}}I(|{\bf{k}}+{\bf{p}}|)$.
To evaluate $I(|{\bf{k}}+{\bf{p}}|)$ we choose
$({\bf{k}}+{\bf{p}})$ along the $y$-axis. Performing the
$\theta_{{\bf{R}}}$ integral which produces $-2\pi i
J_{-1}(|{\bf{k}}+{\bf{p}}|R)$, where $J_{-1}$ is the Bessel
function of the first kind of order $-1$ \cite{Table}, and then
performing the $R$ integral which produces
$\frac{(2\pi)^3i}{|{\bf{k}}+{\bf{p}}|^2}\times\mathcal{O}(1)$, we
find,
\be %&H_{1\rm{P}}=2\pi\Delta_0\sum_{\v k}\sum_{\v
%p}{(k_x-p_x)+i(k_y-p_y)\over |{\bf{k}}+{\bf{p}}|^2 }
%e^{i\theta_{{\bf{k}}+{\bf{p}}}}c^{\dagger}_{{\bf{k}}}c^{\dagger}_{{\bf{p}}}+h.c..
%\nn&=
H_{1\rm{P}}=-(2\pi)^3i\Delta_0\sum_{\v k,\v p}{(k_x+ik_y)^2-(p_x+i
p_y)^2\over |{\bf{k}}+{\bf{p}}|^3 }
c^{\dagger}_{{\bf{k}}}c^{\dagger}_{{\bf{p}}}+{\rm{h.c.}}\nonumber
\ee
%In going from the first to the second line of \Eq{p2} we have
%used the fact  $e^{i\theta_{{\bf k}+{\bf p}}}=\frac{1}{|{\bf k}+{\bf
%p}|}[(k_x+p_x)+i(k_y+p_y)]$.
Finally, using angular momentum expansion of the fermion operators
and noting that a function of $|{\bf k}+{\bf p}|$ is periodic in
$(\theta_{{\bf k}}-\theta_{{\bf p}})$ and hence can be Fourier
expanded as $$ \frac{1}{|{\bf k}+{\bf
p}|^3}=\sum_{m}u_m(k,p)e^{-im(\theta_{{\bf k}}-\theta_{{\bf
p}})},$$ we rewrite $H_{1\rm{P}}$ as, \be
&H_{1\rm{P}}=-(2\pi)^2i\Delta_0\sum_{{\bf k},{\bf
p}}\sum_{m,m_1,m_2}\Big(\frac{k^2e^{i2\theta_{{\bf
k}}}-p^2e^{i2\theta_{{\bf p}}}}{\sqrt{kp}}\Big)\nn&\times u_m(k,p)
e^{-i(m_1+m)\theta_{{\bf k}}}e^{-i(m_2-m)\theta_{\v
p}}c^{\dagger}_{m_1k}c^{\dagger}_{m_2p}+\rm{h.c.}.\nn& \label{p3}
\ee  Performing the $\theta_{{\bf k}}$ and $\theta_{{\bf p}}$
integrals in \Eq{p3} for the two different terms, we get the
conditions $$m_1=2-m,~~m_2=m,~{\rm and}~m_1=-m,~~m_2=2+m$$
respectively. Since $\frac{1}{|{\bf k}+{\bf p}|^3}$ depends on
$\cos(\theta_{\v k}-\theta_{\v p})$, and hence is even in
$(\theta_{{\bf k}}-\theta_{{\bf p}})$, $u_m(k,p)=u_{-m}(k,p)$. In
addition, since ${1\over |\v k+\v p|^3}$ is symmetric in the $\v
k\leftrightarrow\v p$ exchange, $u_m(k,p)=u_m(p,k)$. Using these,
we perform the transformation $m \rightarrow -m$ and interchange
$k$ and $p$ in the second term of \Eq{p3} to arrive at the simple
form for the pairing term,
\begin{equation}
H_{\rm{1P}}=-2i\Delta_0\sum_m\int dk dp
~k^2\sqrt{kp}~u_m(k,p)c^{\dagger}_{2-m,k}c^{\dagger}_{m,p} +
\rm{h.c}. \label{Eq:Pairingvortex4}
\end{equation}
\\

In the above pairing Hamiltonian, each angular momentum channel $m$
is coupled to channel ($2-m$). The only exception is the channel
$m=1$ which is decoupled from the rest. In this channel, writing
$k=k_{\rm{F}}+q$ and $p=k_{\rm{F}}+q^{\prime}$ for small $q$ and
$q^{\prime}$, and noting that the coefficient of
$c^{\dagger}_{1,q}c^{\dagger}_{1,q^{\prime}}$ must be odd in
($q-q^{\prime}$) from the fermion anticommutation relation
(otherwise, the integrals over $q$ and $q^{\prime}$ in the pairing
term below will give zero since the fermion bilinear changes sign
under $q \leftrightarrow q^{\prime}$), the Hamiltonian for the $m=1$
channel takes the form, \be
H_1&=&\frac{1}{(2\pi)^2}\int_{-\Lambda}^{\Lambda}dq~v_{\rm{F}}q~c^{\dagger}_{1,q}c_{1,q}+
i\Delta_0\int_{-\Lambda}^{\Lambda}dqdq^{\prime}
A(q-q^{\prime})\nn&\times&
c^{\dagger}_{1,q}c^{\dagger}_{1,q^{\prime}} +\rm{h.c.}\nonumber\\
\label{fc1} \ee where $A(q-q^{\prime})$ is an odd function of
$(q-q^{\prime})$, which, in leading order, is simply proportional to
$(q-q^{\prime})$. Defining a two component fermion operator $
\chi^\dagger_q=(c^\dagger_{1,q},c_{1,q}),$  $H_1$ becomes \Eq{h}
where $\chi^{\dagger}(x)=(c^{\dagger}(x),c(-x))$ is the Fourier
transform of $\chi_q$ and $m(x)$ is the Fourier transform of
$i\Delta_0 A(q)$. Due to the odd nature of $A(q)$, the mass term
satisfies $m(-x)=-m(x)$.
\\

Now we prove the index theorem for \Eq{h}. By writing the
quasiparticle operator as $$ \gamma^{\dagger}=\int
dx[\eta_1(x)\chi^\dagger_1(x) +\eta_2(x)\chi^\dagger_2(x)]$$ where
$\chi^\dagger_1(x)=c^{\dagger}(x)$ and $\chi^\dagger_2(x)=c(-x)$ and
demanding that $[H,\gamma^{\dagger}]=\e \gamma^{\dagger}$, we obtain
the following equation for $\eta^{\rm{T}}(x)=(\eta_1(x),\eta_2(x))$,
\be -iv_F\s_z\p_x\eta(x)+m(x)\s_x\eta(x)={\e\over
2}\eta(x).\label{mf}\ee The rest of the proof for the zero mode is
identical to that in the Dirac fermion case. To see that the zero
energy quasiparticle is a Majorana fermion we recall that the zero
energy solution of \Eq{mf} is an eigenstate of $\s_y$. Explicitly it
takes the form %$ \eta^{\rm{T}}(x)=e^{{\lambda\over v_F}\int_0^x
%m(y)dy}~~{e^{i\pi/4}\over\sqrt{2}}\begin{pmatrix}1,
%-i\end{pmatrix}.$
$$ \eta(x)=e^{{\lambda\over v_F}\int_0^x
m(y)dy}~~{e^{i\pi/4}\over\sqrt{2}}\begin{pmatrix}1\cr
-i\end{pmatrix}.$$ Here we have chosen a particular global phase for
the $\s_y$ eigenvector. Like the Dirac fermion zero modes, the value
of $\lambda$ here is $\pm 1$ when the mass profile satisfies $
m(x)=\mp{\rm sign(x)}|m(x)|$. The corresponding quasiparticle
operator takes the form,
\begin{equation} \gamma^{\dagger}={1\over{\cal N}}\int
dx~e^{{\lambda\over v_F}\int_0^x m(y)dy}~
\frac{e^{i\pi/4}}{\sqrt{2}}\Big(c_1(x)-ic^{\dagger}_1(-x)\Big),
\end{equation}
where ${\cal N}$ is a normalization factor. Since $e^{{\lambda\over
v_F}\int_0^x m(y)dy}$ is an even function of $x$ (since $m(x)$ is an
odd function),  one can easily verify that
$\gamma^{\dagger}=\gamma$, i.e., the quasiparticle is a Majorana
fermion.

In angular momentum channels other than $m=1$, channel $m$ is
coupled to channel ($2-m$). For these channels, the Hamltonian is of
the form \be
H_m&=&\frac{1}{(2\pi)^2}\int_{-\Lambda}^{\Lambda}dq~v_{\rm{F}}
q~(c^{\dagger}_{mq}c_{mq}+c^{\dagger}_{2-mq}c_{2-mq})\nn
&+&i\Delta_0\int_{-\Lambda}^{\Lambda}dqdq^{\prime} B(q,q^{\prime})
c^{\dagger}_{mq}c^{\dagger}_{2-mq^{\prime}} +\rm{h.c.}.\nonumber \ee
In this case there is no requirement that $B(q,q')$ has to change
sign upon $q\leftrightarrow q'$ since the fermion operators in the
pairing term are from two different angular momentum channels. Thus,
generically, $B(0,0)$ is non-zero, and the spectrum is gapped. In
this case the real space Hamiltonian resembles \Eq{h2} with $
\psi^{\dagger}(x)$ the Fourier transform of
$(c_{mq},c^{\dagger}_{2-mq})$, and the mass term does not change
sign.  Hence, there is exactly one zero mode in the spectrum.
\\

For the system with a winding-number-two vortex, the factor
$e^{i\theta_{{\bf R}}}$ in \Eq{Pairingvortex} is replaced by
$e^{2i\theta_{{\bf R}}}$. This simple modification, as we show
below, gets rid of the uncoupled angular momentum channel.
Consequently, like the $m\ne 1$ angular momentum channels in the
winding-number-one vortex discussed above, the real space
Hamiltonian is of the form in \Eq{h2} where the mass term does not
change sign. Therefore the Jackiw-Rebbi theorem does not apply and
there is no zero energy mode in the spectrum.

With the modification $e^{i\theta_{{\bf R}}}\ra e^{2i\theta_{{\bf
R}}}$ the function $I(\v k+\v p)$ in Eq.(\ref{I})  now satisfies,
$
I(R_{\theta}({\bf{k}}+{\bf{p}}))=e^{2i\theta}I({\bf{k}}+{\bf{p}}).$
It follows that, $
I({\bf{k}}+{\bf{p}})=e^{2i\theta_{{\bf{k}}+{\bf{p}}}}I(|{\bf{k}}+{\bf{p}}|).$
Doing similar calculation as earlier  we obtain, \be &H_{2\rm{P}}=
-(2\pi)^2\Delta_0\sum_{{\bf k},{\bf p}}\frac{r(k,p,\theta_{{\bf
k}},\theta_{{\bf p}})}{\sqrt{kp}}\sum_{m,m_1,m_2}
v_m(k,p)\nn&\times e^{-i(m_1+m)\theta_{\v k}}
e^{-i(m_2-m)\theta_{\v p}}
c^{\dagger}_{m_1k}c^{\dagger}_{m_2p}+h.c., \nonumber \ee where \be
r(k,p,\theta_{{\bf k}},\theta_{{\bf
p}})&=&\Big(k^3e^{3i\theta_{{\bf k}}}+k^2pe^{2i\theta_{{\bf
k}}}e^{i\theta_{{\bf p}}}-p^2ke^{2i\theta_{{\bf
p}}}e^{i\theta_{{\bf k}}}\nn&-&p^3e^{3i\theta_{{\bf
p}}}\Big).\nonumber\ee and $v_m(k,p)$ is the Fourier component of
$\frac{1}{|{\bf k}+{\bf p}|^4}$. Finally, performing the
$\theta_{{\bf k}}$ and $\theta_{{\bf p}}$ integrals, and
transforming $m\rightarrow -m$ along with interchanging $k$ and
$p$ in the third and fourth terms of $H_{2\rm{P}}$, we arrive at
the equation analogous to Eq.(\ref{Eq:Pairingvortex4}),
\begin{eqnarray}
H_{2\rm{P}}&=&2\Delta_0\sum_m\int dkdp~ (k^3+k^2p)\sqrt{kp}~
v_m(k,p)\nonumber\\&\times&c^{\dagger}_{3-mk}c^{\dagger}_{mp} +
\rm{h.c}.\nn \label{Eq:Pairingdoublevortex2}\nonumber
\end{eqnarray}
It is clear from this equation that, in the case of the vortex
with winding number two, no angular momentum channel in the
pairing term decouples from the rest.

We can generalize the above calculations to the cases of arbitrary
odd and even winding number vortices. For a vortex with an odd
winding number, $2n-1$, because of the factor
$e^{i(2n-1)\theta_{{\bf R}}}$ in the pairing term, angular
momentum channel $m$ is coupled to channel ($2n-m$). In this case,
there is always a channel, $m=n$, which is decoupled from the
rest. In this channel, quite generally, the Hamiltonian maps on
\Eq{h} with a mass term which is an odd function simply because of
the fermion anticommutation relation. The index theorem we proved
in this paper then implies the existence of a zero energy Majorana
fermion quasiparticle. In contrast, for a vortex with an even
winding number, $2n$, angular momentum channel $m$ is coupled to
channel ($2n+1-m$) by the pairing term. Consequently, like in the
case of the vortex with winding number two, there is no decoupled
angular momentum channel and there is no zero mode. Note that in
the physics discussed above the breaking of time reversal symmetry
is crucial, since, for 2D non-chiral $p$-wave superconductors, the
bulk
itself is gapless. \\

For $s$-wave superconductors, pairing occurs between the fermions
with opposite spins. In this case, for a vortex with winding number
one, one can show that the pairing term couples angular momenta $m$
and $1-m$, like in the uniform case discussed above. So there is no
isolated channel. For a vortex with winding number two, the channel
$m=1$ decouples, and the Hamiltonian in this channel takes the form
\be
H_2&=&\frac{1}{(2\pi)^2}\sum_\s\int_{-\Lambda}^{\Lambda}dq~v_{\rm{F}}q~c^{\dagger}_{1q\s}c_{1q\s}+
\Delta_0\int_{-\Lambda}^{\Lambda}dqdq^{\prime}
C(q,q^{\prime})\nn&\times&
c^{\dagger}_{1q\ua}c^{\dagger}_{1,q^{\prime}\da}
+\rm{h.c.}.\label{s-wave}
 \ee However, in Eq.~\ref{s-wave}, because of the opposite spins of the
 fermions in the pairing term, the
anticommutation relations fail to produce an odd mass term. Thus,
there is no zero mode at the vortices of an $s$-wave superconductor
\cite{Caroli}.

To conclude, we have proven an index theorem for the existence of
zero modes in the vortex cores of chiral $p$-wave superconductors.
The theorem, which is an analog of that in 1D Dirac fermion field
theory with a mass soliton, correctly predicts a zero energy
Majorana quasiparticle at the core of a vortex with odd winding
number, and no such zero mode for a vortex with even winding number.
The formalism also correctly captures the low energy quasiparticle
physics at the vortex cores of $s$-wave superconductors.

We thank A. Seidel, Y. B. Kim and A. Vishwanath for interesting
discussions. S.T. thanks the Physics Department, University of
California, Berkeley, for hospitality. This research is supported by
the DOE Grant No. DE-AC03-76SF00098, and ARO-DTO, ARO-LPS, and NSF.

\end{document}